\documentclass[usegraphicx,usenatbib]{mn2e}

\usepackage{url}
\usepackage[breaklinks=true]{hyperref} 
\usepackage{twoopt}
\usepackage[english]{babel}          

\usepackage{natbib}
\bibpunct{(}{)}{;}{a}{}{,} 

\usepackage[utf8]{inputenc}
\usepackage[normalem]{ulem}
\usepackage{siunitx}
\usepackage{amsmath, amssymb}
\usepackage[T1]{fontenc} 
\usepackage{aecompl} 

\usepackage{multirow,bigstrut,ctable}
\usepackage[normalem]{ulem}
\usepackage{rotating}
\usepackage[varg]{txfonts} 
\usepackage{threeparttable}

\def \deg         {\text{$^{\circ}$}}
\def \arcmin      {\text{$^\prime$}}
\def \arcsec      {\text{$^{\prime\prime}$}}

\def \jybeam     {Jy\,beam$^{-1}$}
\def \mjybeam     {mJy\,beam$^{-1}$}

\newcommand{\ion}[2]{#1$\;${\small\rmfamily\@Roman{#2}}\relax}
\newcommand{\Hii}{\text{H\textsc{ii}}}
\newcommand{\Hi}{\text{H\textsc{i}}}

\title[Radio spectral index map and catalogue]
      {A radio spectral index map and catalogue at 147--1400 MHz covering 80~per cent of the sky}

\author[F.~de~Gasperin et~al.]{F. de Gasperin$^{1}$, H. T. Intema$^{1}$, D. A. Frail$^{2}$\\
$^{1}$ Leiden Observatory, Leiden University, P.O.Box 9513, NL-2300 RA, Leiden, The Netherlands\\
$^{2}$ National Radio Astronomy Observatory, 1003 Lopezville Road, Socorro, NM 87801-0387, USA}
\begin{document}

\date{}
\pagerange{\pageref{firstpage}--\pageref{lastpage}} \pubyear{2017}
\maketitle

\label{firstpage}

\begin{abstract}
The radio spectral index is a powerful probe for classifying cosmic radio sources and understanding the origin of the radio emission.
Combining data at 147~MHz and 1.4~GHz from the TIFR GMRT Sky Survey (TGSS) and the NRAO VLA Sky Survey (NVSS), we produced a large-area radio spectral index map of $\sim 80$~per cent of the sky (Dec~$>-40\deg$), as well as a radio spectral index catalogue containing 1\,396\,515 sources, of which 503\,647 are not upper or lower limits. Almost every TGSS source has a detected counterpart, while this is true only for 36 per cent of NVSS sources. We released both the map and the catalogue to the astronomical community.
The catalogue is analysed to discover systematic behaviours in the cosmic radio population. We find a differential spectral behaviour between faint and bright sources as well as between compact and extended sources. These trends are explained in terms of radio galaxy evolution. We also confirm earlier reports of an excess of steep-spectrum sources along the galactic plane. This corresponds to 86 compact and steep-spectrum source in excess compared to expectations. The properties of this excess are consistent with normal non-recycled pulsars, which may have been missed by pulsation searches due to larger than average scattering along the line of sight.
\end{abstract}


\begin{table*}
\centering
\begin{threeparttable}
\begin{tabular}{lll}
Column name\tnote{a} & Format & Notes\tnote{b}\\
\hline
RA & degree & S,M,C,I: weighted average of NVSS and TGSS detections \bigstrut[t]\\
DEC & degree & S,M,C,I: weighted average of NVSS and TGSS detections \\
Total\_flux\_NVSS & Jy & I: sum of detection fluxes; U: set to 0 \\
E\_Total\_flux\_NVSS & Jy & I: quadratic mean of detection errors; U: set to 0 \\
Peak\_flux\_NVSS & \jybeam & I: maximum of detection peak fluxes; U: set to 0 \\
E\_Peak\_flux\_NVSS & \jybeam & I: error relative to the maximum of detection peak fluxes; U: set to 0 \\
Rms\_NVSS & \jybeam & Local rms noise \\
Total\_flux\_TGSS & Jy & I: sum of detection fluxes; L: set to 0 \\
E\_Total\_flux\_TGSS & Jy & I: quadratic mean of detection error; L: set to 0 \\
Peak\_flux\_TGSS & \jybeam & I: maximum of detection peak fluxes; L: set to 0 \\
E\_Peak\_flux\_TGSS & \jybeam & I: error relative to the maximum of detection peak fluxes; L: set to 0 \\
Rms\_TGSS & \jybeam & Local rms noise \\
Spidx & -- & Spectral index calculated as described in the text \\
E\_Spidx & -- & L,U: set to 0 \\
s2n & -- & Peak\_flux over Island Rms, see text for details \\
S\_code & S,M,C,L,U,I & Source code as described in the text \\
Num\_match & -- & Number of matched sources in the same island \\
Num\_unmatch\_NVSS & -- & Number of NVSS unmatched sources in the same island \\
Num\_unmatch\_TGSS & -- & Number of TGSS unmatched sources in the same island \\
Isl\_id & -- & Island identification number \\
Source\_id & -- & Unique source identification number \\
\end{tabular}
\begin{tablenotes}
    \item[a] An ``E\_'' in front of the column name denotes the error of that quantity.
    \item[b] Some notes are valid only for specific entry types as saved in the S\_code column.
\end{tablenotes}
\end{threeparttable}
\caption{Description of catalogue's columns. The full catalogue is available as online supplementary material and at \url{http://tgssadr.strw.leidenuniv.nl/spidx}.}\label{tab:catalogue}
\end{table*}

\section{Introduction}

Our understanding of many physical phenomena producing cosmic radio emission is linked to the spectral energy distributions (SED) characteristics of the radiation. Radio SED are usually smooth and can be described with a power law. Therefore, one can define a relevant quantity, called \textit{spectral index}, as the measure of the radio SED slope in a log-log plane. For a frequency $\nu$ and a flux density $S_\nu$, the spectral index $\alpha$ is given by
\begin{equation}\label{eq:spidx}
 S_\nu \propto \nu^{\alpha} \rightarrow \alpha\left( \nu \right) = \frac{\partial \log S_\nu}{\partial \log \nu}\,.
\end{equation}
Radio SEDs often decrease with increasing frequency, in which case $\alpha$ is a negative number. This made the opposite sign convention for $\alpha$ ($S_\nu \propto \nu^{-\alpha}$) quite common in the literature. In this paper we will use the convention of Eq.~\ref{eq:spidx}.

The spectral index of compact radio sources and spatial variations across extended sources is a fundamental tool in the study of all radio source populations including galaxies, active galactic nuclei (AGN), clusters of galaxies, pulsars, \Hii{} regions, and supernova remnants. The spectral index is important in disentangling different mechanisms of radio production in galaxies \citep{Lisenfeld2000}, which generate a flat-spectrum component, due to thermal Bremsstrahlung from H\textsc{ii} regions, and a steep-spectrum component, due to the synchrotron emission from cosmic ray electrons. In the case of radio galaxies' lobes, the SED generally shows a negative curvature, with the spectrum becoming steeper at high frequency \citep{Konar2006, Jamrozy2008, McKean2016}. The most important mechanisms here are synchrotron and inverse Compton losses at high frequencies and synchrotron self-absorption at low frequencies. The SED study in this case gives insights on the source age, energetics and plasma composition. AGN cores show a flat spectrum that is thought to be the result of self-absorption rather than a flat electron energy distribution. In galaxy clusters, we find haloes, relics and other radio sources with complex morphologies. A key part of source classification comes from the spectral index information \citep{vanWeeren2010a, deGasperin2017c}. Supernova remnant studies also benefit from spectral index information to investigate particle acceleration in shocks and their interaction with the environment \citep{Anderson1993, Moffett1994, Castelletti2011}.

A fast procedure to create a spectral index catalogue is to cross-match the source catalogues produced by two or more surveys at different frequencies \citep{Vollmer2004a,ki08,Tiwari2016}. However, this approach has some important limitations: (i) the two catalogues were likely produced with different techniques and are subject to different (uncontrolled) biases, (ii) if the resolution of the two surveys is not equal (as it is almost always the case), some sources in the low-resolution catalogue could match multiple sources in the high resolution one in a non-trivial way, (iii) extended sources are usually treated as collections of gaussians whose matching from catalogues is complicated.

\begin{figure}
\centering
\includegraphics[width=.5\textwidth]{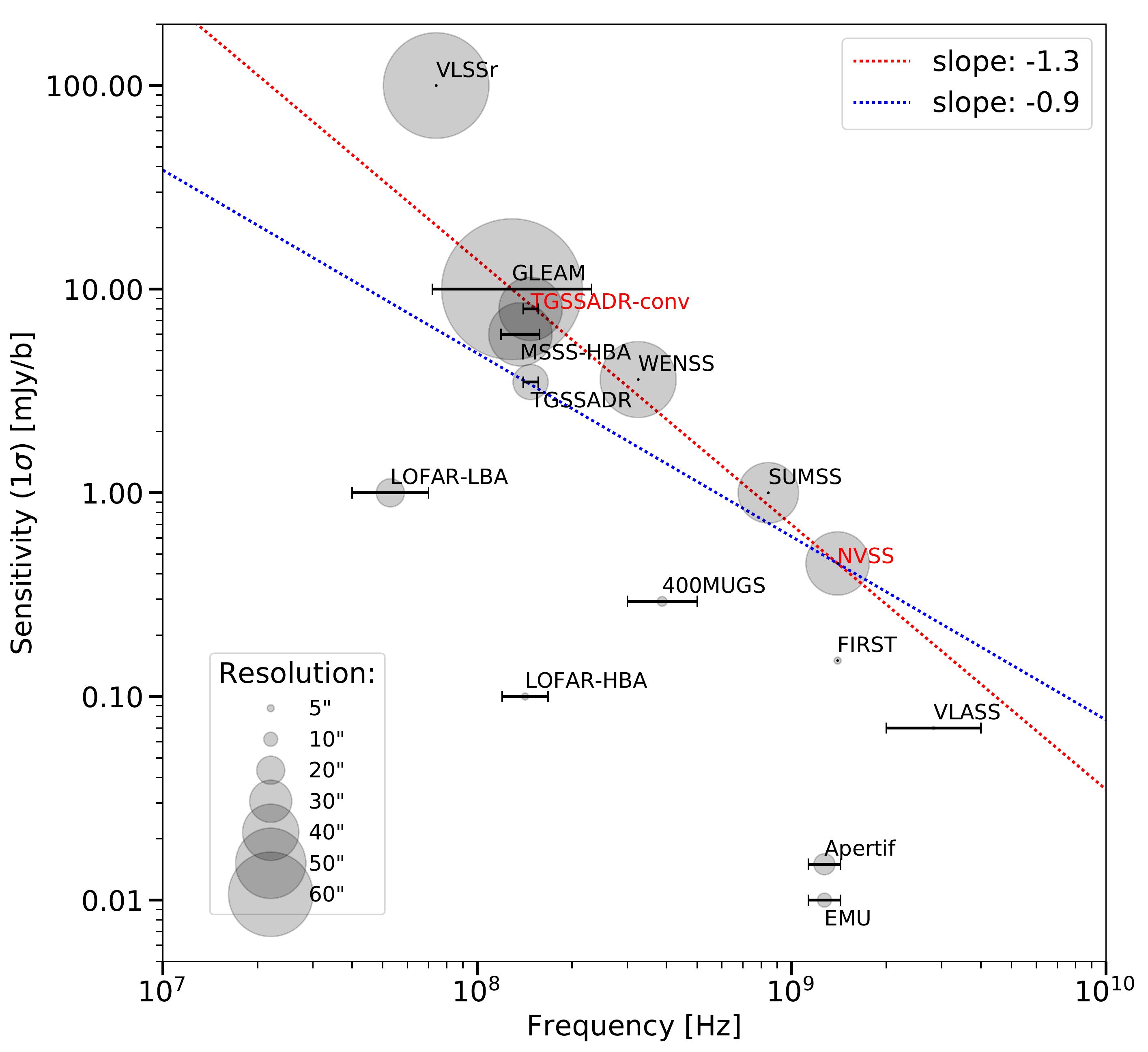}
\caption{Past and planned large radio surveys. The grey circle diameters are proportional to the survey beam size as shown in the bottom left corner. The two surveys used for this work are highlighted in red. For wide band surveys we show the frequency coverage using horizontal lines. NVSS and TGSS (convolved at 45\arcsec) are equally sensitive to sources with a spectral index $\alpha=-1.3$. Apart from FIRST, all surveys below the blue line are currently planned or on-going. References: GLEAM \citep[GaLactic and Extragalactic All-sky Murchison Widefield Array survey][]{Hurley-Walker2017}; MSSS-HBA \citep[LOFAR Multifrequency Snapshot Sky Survey][]{Heald2015}; TGSS ADR1 \citep[TIFR GMRT Sky Survey - Alternative Data Release 1][]{Intema2017}; VLSSr \citep[VLA Low-frequency Sky Survey redux][]{Lane2014}; FIRST \citep[Faint Images of the Radio Sky at Twenty Centimetres][]{Becker1995}; NVSS \citep[1.4 GHz NRAO VLA Sky Survey][]{Condon1998}; WENSS \citep[The Westerbork Northern Sky Survey][]{Rengelink1997}; SUMSS \citep[Sydney University Molonglo Sky Survey][]{Bock1999a}; EMU \citep[Evolutionary Map of the Universe][]{Norris2011}; Apertif \citep[][]{Rottgering2012}; VLASS (VLA Sky Survey); LOFAR-HBA \citep[LOFAR High Band Antenna][]{Shimwell2016a}; LOFAR-LBA (LOFAR Low Band Antenna; de Gasperin et al. in prep.)}\label{fig:comp_sensitivity}
\end{figure}

In this paper we produced a spectral index map and catalogue by matching specially re-processed images from two radio surveys at different frequencies: NVSS and TGSS. These surveys were chosen because they are the most sensitive metre and centimetre wavelength surveys currently available that cover a large fraction of the sky, and their order-of-magnitude separation in frequency provides a large lever arm for precise spectral index calculation. The surveys also have comparable angular resolution and similar rms noise sensitivity for sources with $\alpha=-0.9$ (or $\alpha=-1.3$ for matched-resolution TGSS data, see Fig.~\ref{fig:comp_sensitivity}). The resulting spectral index catalogue includes $\sim 1.4$ million entries, making it the largest collection of spectral index information to date. Furthermore, we provide the community with a spectral index map of 80~per cent of the sky that can be used to classify radio sources.

In Sec.~\ref{sec:catalogue} of this paper we present a spectral index catalogue of the radio sky north of Declination ($\delta$) $-40$\deg{}, obtained using the radio surveys NVSS and TGSS. In Sec.~\ref{sec:map} present a pixel-by-pixel masked spectral index map of the same region of the sky. In Sec.~\ref{sec:discussion} we investigate two outstanding questions in extra-galactic and galactic research to illustrate the capabilities of this new catalogue. We end in Sec.~\ref{sec:conclude} with some suggested future applications of the spectral index catalogue and the maps.

\section{Spectral index catalogue}
\label{sec:catalogue}

\begin{figure*}
\centering
\includegraphics[width=\textwidth]{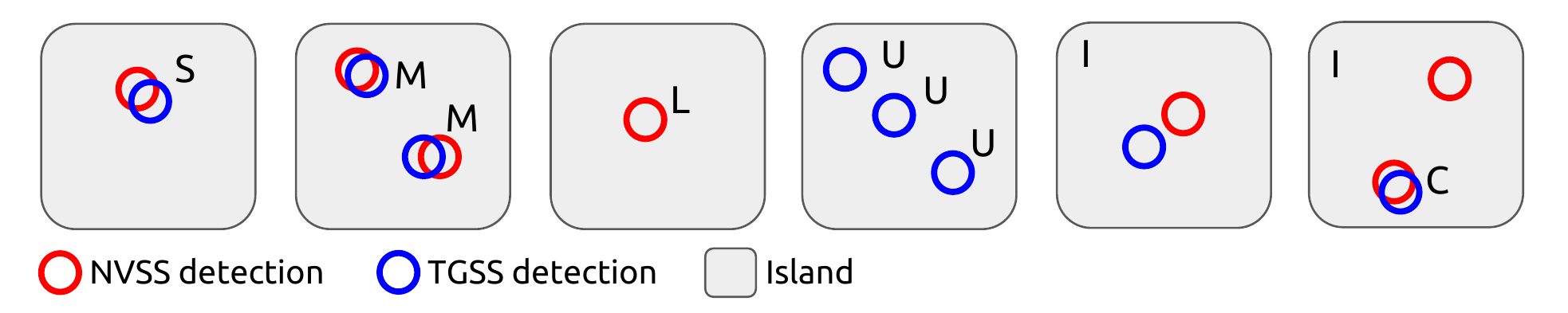}
\caption{Some classification examples. Coloured circles represent the locations of an NVSS/TGSS detections while grey areas represent islands of contiguous emission as described in the text. When two circles overlap they are ``matched'': their centres are closer than 15\arcsec. See Sec.~\ref{sec:catalogue} for more details.}\label{fig:cartoon}
\end{figure*}

Both the spectral index map and catalogue were extracted from the following radio surveys:
\\

\textbf{NVSS} -- The NRAO VLA Sky Survey \citep{Condon1998} covers the sky north of $\delta = -40\deg$ (82~per cent of the celestial sphere). The survey was made with the Very Large Array (VLA) in D and DnC configurations, at 1.4 GHz in full polarisation. However, for this work we used only Stokes I images. The resolution is $45\arcsec$ and the background rms noise is nearly uniform at $0.45$~\mjybeam{}. The overall astrometric accuracy is better than 1\arcsec{} in RA and Dec. Due to the compactness of the VLA configuration used, the surface brightness of extended sources is fairly well reconstructed up to scales of $\sim 16$\arcmin. At the same time, extended and un-modelled surface brightness from the galactic plane lower the fidelity of images at low galactic latitude.
\\

\textbf{TGSS ADR1\footnote{For the rest of the paper we will use TGSS as an abbreviation for TGSS ADR1}} -- The TIFR GMRT Sky Survey Alternative Data Release 1 \citep{Intema2017} covers the full sky visible from the Giant Meterwave Radio Telescope (GMRT), i.e. north of $\delta = -53\deg$ (90~per cent of the celestial sphere). The survey is observed at a frequency of 147 MHz with a resolution of $25\arcsec$ and a median rms noise of $3.5$~\mjybeam{}. The overall astrometric accuracy is better than 2\arcsec{} in RA and Dec, while the flux density accuracy is estimated to be $\sim10$~per cent for most of the survey area. The higher resolution of GMRT, combined with the data reduction strategy that down-weights the short baselines, reduced both the sensitivity of TGSS to extended emission as well as the presence of artefacts along the galactic plane due to bright, extended sources. The largest detectable angular scale in TGSS is of order a few arcmin.
\\

To mitigate many of the problems of blind catalogue matching, we re-processed and combined the images of the NVSS and TGSS surveys. To match the lower NVSS resolution, we convolved the TGSS images to a 45\arcsec{} circular beam. As a result, the median rms noise went up to $8$~\mjybeam{}, which is 2.2 times higher than the original TGSS. Next, we combined and regridded the TGSS images to exactly match the NVSS mosaic images. Blank pixels in one survey were blanked in the other (and vice versa), such as the hole of $\sim100$ square degrees in the TGSS Northern hemisphere, and the NVSS southern sky below $\delta=-40\deg$. We obtained two sets of $2325$ images with a matched resolution, pixel grid, and coverage, which were the basis for further processing.

We ran the source finder PyBDSF\footnote{Formerly named PyBDSM} \citep{Mohan2015} on both sets of survey images with the same parameters. This tool uses a running rms box to calculate the local rms background ($\sigma_l$) across each image. Island masks of contiguous emission are created by selecting pixels with values $>4\sigma_l$ and expanding them locally using only pixels with values $>3\sigma_l$. The surface brightness of each pixel island is fitted with one or more Gaussians. However, the number, exact position, and shape of these Gaussians might differ in the two surveys as they depend on the detailed structure of the source and, to certain extent, on the local noise. Therefore, close-by Gaussians are grouped into sources that will be used for the cross-match. For each pair of images, the outputs of PyBDSF were two rms maps, two island masks, and two source catalogues.

Per image pair we combined the output of the two PyBDSF runs. We combined the two island masks to obtain a global mask indicating where at least one survey had a detection. This is a Boolean OR done pixel-by-pixel. We also cross-matched the two catalogues to identify the closest detection with a maximum separation of 15\arcsec{} (a third of the synthesized beam size). By construction, each detection, even those with no counterpart, must lie within one island. At this point sources were classified into separate categories:
\begin{itemize}
\item single (S): a matched source (detection in both NVSS and TGSS) with no other detections in the same island -- these sources are simple in both surveys, usually point-like or slightly extended;
\item multiple (M): a matched source with other matched sources (but no unmatched sources) in the same island -- these sources are mostly double-lobe radio galaxies or collections of nearby point sources;
\item lower limit (L): a detection in NVSS with no TGSS detection in the same island -- multiple NVSS detections in the island create multiple L-entries in the catalogue;
\item upper limit (U): a detection in TGSS with no NVSS detection in the same island -- multiple TGSS detections in the island create multiple U-entries in the catalogue;
\item complex (C): a matched source with unmatched detections in one of the two surveys in the same island -- these sources might be part of a more complex object, e.g. a lobe of a radio galaxy where the other lobe is too faint to be detected in one of the surveys;
\item island (I): if an island is not only made by matched sources (S, M) or unmatched source of the same category (L, U), the global values of that island are saved as a separate entry (with intrinsically poorly defined positions) -- complex sources (C) are always part of an island source (I).
\end{itemize}
Fig.~\ref{fig:cartoon} shows an example diagram for each source category as defined above. From here onwards, we define \emph{full detections} as sources having the code S, M, or C. Upper and lower limits with peak flux to local rms noise ratio (s2n -- also saved as a column in the catalogue) below 5 are removed from the final catalogue. For full detections and islands this criterion is relaxed. Since those sources have detections with a confidence $>4\sigma_l$ in both surveys, each simultaneous detection is $>5\sigma$ significant. In this way we exploit the combined detection of faint sources in both surveys to increase the catalogue completeness.

For upper and lower limits the spectral index is calculated using an upper limit of $3\sigma_l$ in the survey with no detection (no spectra index error is estimated). For S, M, C, and I sources the spectral indices are calculated by bootstrapping the flux densities 1000 times from both detections, assuming that the flux density errors estimated by the source finder are the 1-sigma uncertainties of a Gaussian distribution. We then used the mean and standard deviation of the resulting distribution of spectral indices as spectral index estimator and relative error. We note that the source finder can sometimes underestimate flux density errors, which in turn would provide an underestimated spectral index error value. Coordinates are calculated by making a weighted average of the coordinates of all sources involved in the catalogue entry. Each catalogue entry has listed the number of upper/lower limits and matched sources in the same island of that detection. Furthermore, the island identification (Isl\_id) associated with the detection is also reported. This value can be used to group detections of the same island. The catalogue columns are described in detail in Table~\ref{tab:catalogue}.

The two surveys have different sensitivity to extended emission. Although at 45\arcsec{} resolution most of the radio sky is unresolved, this might bias some results for nearby or very large sources. However, if the extension of a source is visible only in one survey, our algorithm will most likely fail in associating it to any counterpart. In fact, the source position as determined by PyBDSF would be different in the two surveys and the matching criterion, the two detections closer than 15\arcsec{}, would be not satisfied. As a consequence, the source would be classified as a limit (U/L) or as an island (I).

The process of source matching and classification is repeated for each pair of images, resulting in 2325 spectral index catalogues. Islands that touch an image edge were removed entirely to avoid partially sampled islands. As neighbouring images partially overlap, each source that is closer to the centre of another image than the image used for its detection is removed. This process removes all duplicates from the combined (final) catalogue, retaining only the source closest to an image centre.

\subsection{Properties of the catalogue}
\label{sec:properties}

\begin{figure}
\centering
\includegraphics[width=.5\textwidth]{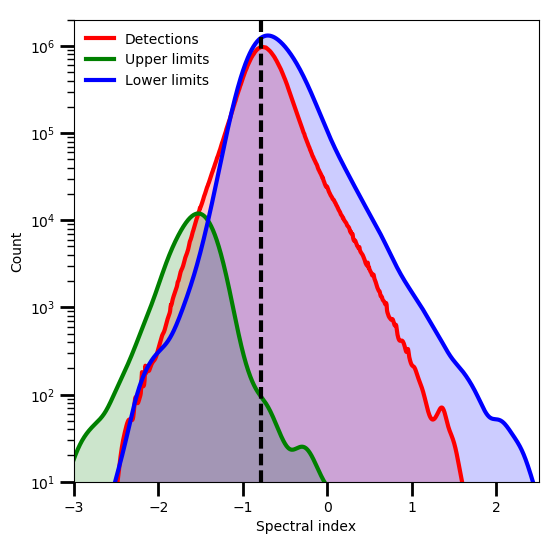}
\caption{Kernel density estimation of the spectral index distribution (each source contributes to the distribution as a normalised Gaussian centred on the spectral index value, with a standard deviation equal to the spectral index error). Sources identified with code S, M, or C are presented as \emph{detections} (red), upper (green) and lower (blue) limits are shown separately. Black dashed line shows the weighted mean spectral index value of the detections ($\alpha=-0.7870 \pm 0.0003,\ \sigma=0.24$).}\label{fig:spidx_kde}
\end{figure}

\begin{figure*}
\centering
\includegraphics[width=\textwidth]{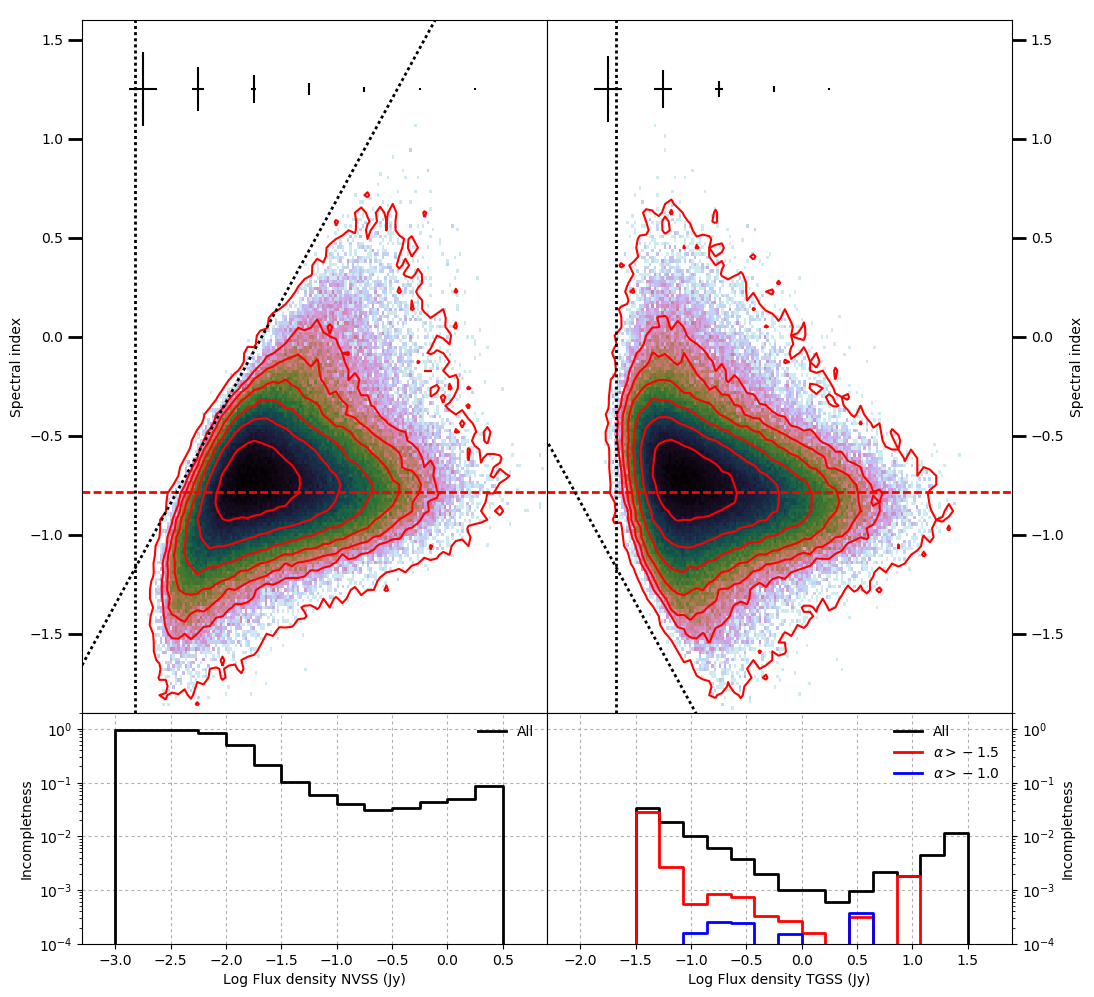}
\caption{Number density of full detections. Contours are at densities of $(5,50,100,200,400,800,1600,3200)$ sources per bin, where the image was sampled with $80\times80$ bins in the ranges: $-3<\log S<1.5$ and $-1.9<\alpha<1.6$. Black dotted lines shows the limits due to three times the median rms noise in both surveys. Sources to the left of those lines went undetected in at least one survey. The mean spectral index and flux density error for several flux density bins are shown at the top of the plot. Red dashed line shows the weighted average spectral index as found in Fig.~\ref{fig:spidx_kde}. Bottom panels show an estimation of the cross-match incompleteness of the catalogue. Black lines show the fraction of the sources that are missing a counterpart per flux density bin. Red and blue line are upper-limits on this fraction and are valid only above the noted spectral index cut.}\label{fig:spidxflux_distrib}
\end{figure*}

The final spectral index catalogue has 1\,396\,515 entries, of which 503\,647 (36~per cent) are full detections, 845\,459 (61 per cent) are lower limits, and 6\,386 ($<1$ per cent) are upper limits. Amongst full detections, 439\,488 sources are ``single'' (S), 35\,202 sources are ``multiple'' (M), and 28\,957 sources are ``complex'' (C). There are 41\,023 (3 per cent) entries marked as ``island'' (I). The high number of lower limits is due to the better sensitivity of NVSS with respect to the TGSS (convolved to 45\arcsec{}) for sources with an average spectral index\footnote{To have the same sensitivity of NVSS for sources with an average spectral index, the convolved TGSS should reach an rms noise of $\sim 2.5$~\mjybeam.}. The distribution of the spectral index values for full detections, upper and lower limits is shown in Fig.~\ref{fig:spidx_kde}. The weighted average spectral index of the full detections is $-0.7870 \pm 0.0003$, with a standard deviation of 0.24.

The spectral index distribution for full detections is not Gaussian nor is it symmetric around the mean. This is due both the intrinsic physical shape of such a distribution and to observational biases. To clarify the effect of the observational biases due to different noise levels in the surveys, in Fig.~\ref{fig:spidxflux_distrib} we plot the number density of full detections as a function of spectral index and flux density. The diagonal dashed lines in both plots (NVSS left and TGSS right) are due to the sensitivity limit in the other survey. While the majority of TGSS sources have an NVSS counterpart, this is not true for NVSS where a strong bias against faint/flat-spectrum sources is induced by the sensitivity limit of TGSS. In general, as expected, faint/flat-spectrum and -- to a lesser extent -- faint/steep-spectrum sources are missing from the catalogue.

\begin{figure}
\centering
\includegraphics[width=.5\textwidth]{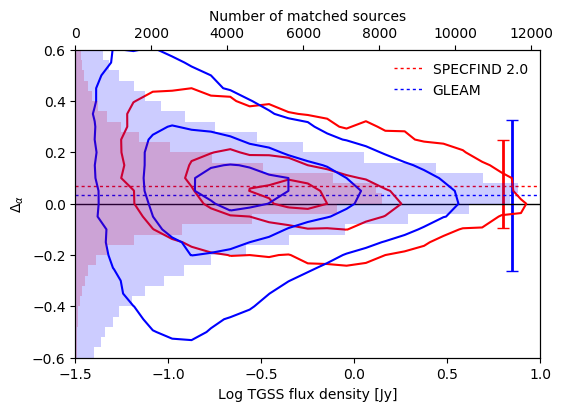}
\caption{Differential spectral index value as a function of TGSS flux density between the catalogue presented in this paper and SPECFIND 2.0 (red) as well as GLEAM (blue). The contours enclose 90, 50 and 10~per cent of all matched sources. The bars on the right show the range in $\Delta_\alpha$ that encloses 90~per cent of all matched sources. The histograms of $\Delta_\alpha$ are also displayed and labelled on the top $x$-axis.}\label{fig:delta}
\end{figure}

We cross-checked our spectral index values against those released by the Galactic and Extragalactic All-sky MWA (GLEAM) survey collaboration \citep{Hurley-Walker2017} and the SPECFIND 2.0 project \citep{Vollmer2010}. The validation has been performed by cross-matching source positions and comparing the spectral index values of matched sources.  GLEAM observed the sky simultaneously from 72 to 231 MHz, which included estimation of the spectral indices over this frequency range. However, we note that GLEAM and our catalogues are not fully independent as GLEAM also uses NVSS to rescale its flux densities. We find an average difference between the spectral index values of $\Delta_\alpha = \left<\alpha_{147\rm\ MHz}^{1400\rm\ MHz} - \alpha_{76\rm\ MHz}^{227\rm\ MHz}\right> = 0.03$. GLEAM spectral indices tend to be on average steeper than those presented in this work. This effect seems to be stronger at lower flux densities. We also cross-checked our catalogue against the multifrequency  SPECFIND 2.0 catalogue, which was obtained by cross-matching a number of survey catalogues (including NVSS) in the frequency range from 149 to 8400 MHz. We also find a mild tendency for our spectral index estimation to be flatter with $\Delta_\alpha = \left<\alpha_{147\rm\ MHz}^{1400\rm\ MHz} - \alpha_{\rm SPECFIND}\right> = 0.07$. About 90~per cent of the spectra lie within $-0.09 < \Delta_\alpha < 0.25$. Results are summarised in Fig.~\ref{fig:delta}.

We identified locations of the sky where the median spectral index is higher/lower than average by tiling the sky using HEALPix \citep[$\rm NSIDE=16$, pixel area: 13 sq deg;][]{Gorski2004}. The standard deviation of the distribution of the median spectral index values is $<0.05$, which corresponds to a combined flux density error of $\sim 12$~per cent in TGSS and NVSS. We suspect that the deviations mostly originate from TGSS, as the spatial variation scales roughly match the TGSS observe strategy  \citep[see][]{Intema2017}. Also, \cite{Hurley-Walker2017a} found that there may be potential systematic errors of about 15~per cent in the TGSS flux density on scales of several degrees, which can systematically bias our spectral index values by $\sim0.06$.

\subsection{NVSS and TGSS completeness and reliability}
\begin{figure}
\centering
\includegraphics[width=.5\textwidth]{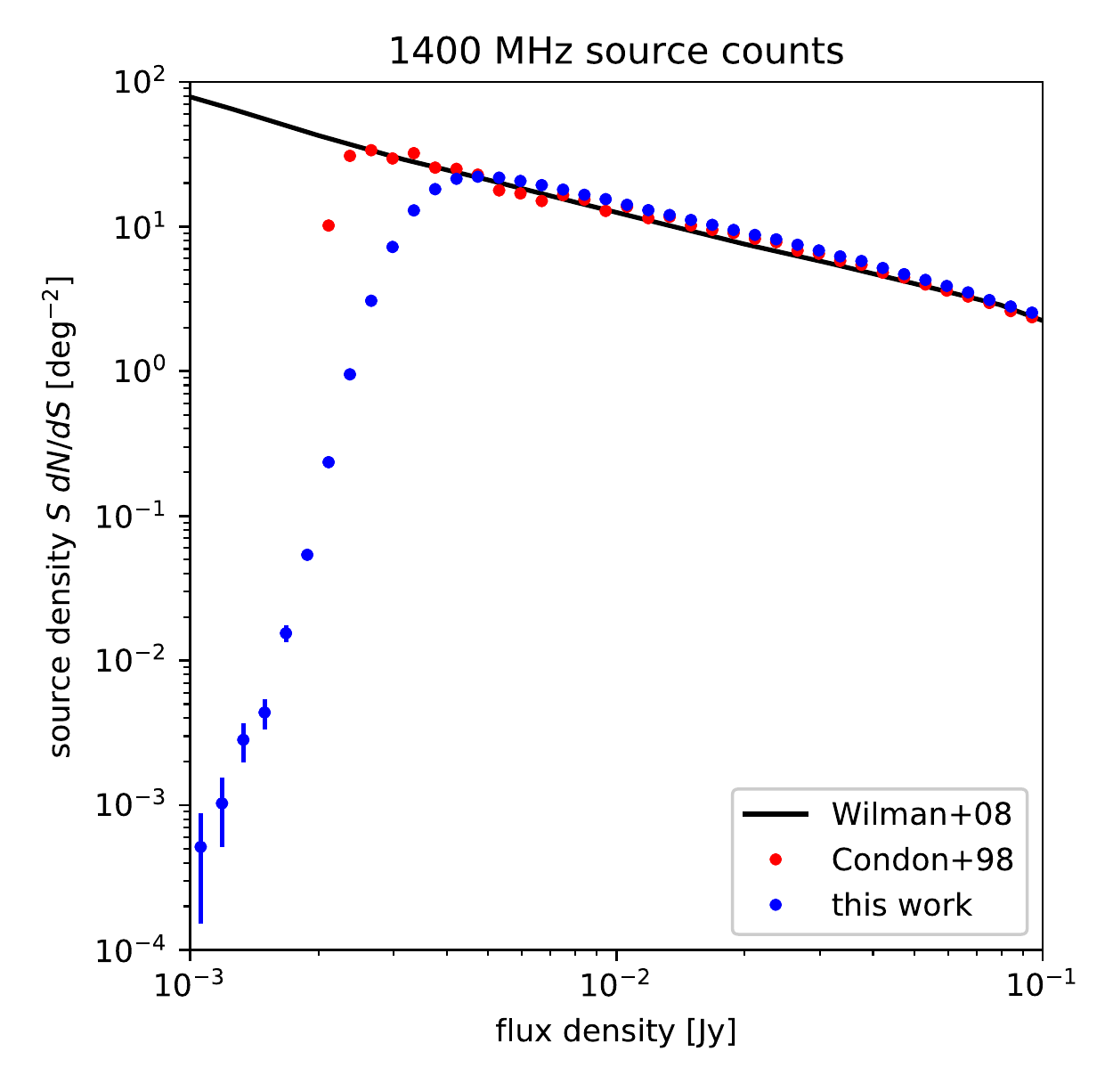}
\includegraphics[width=.5\textwidth]{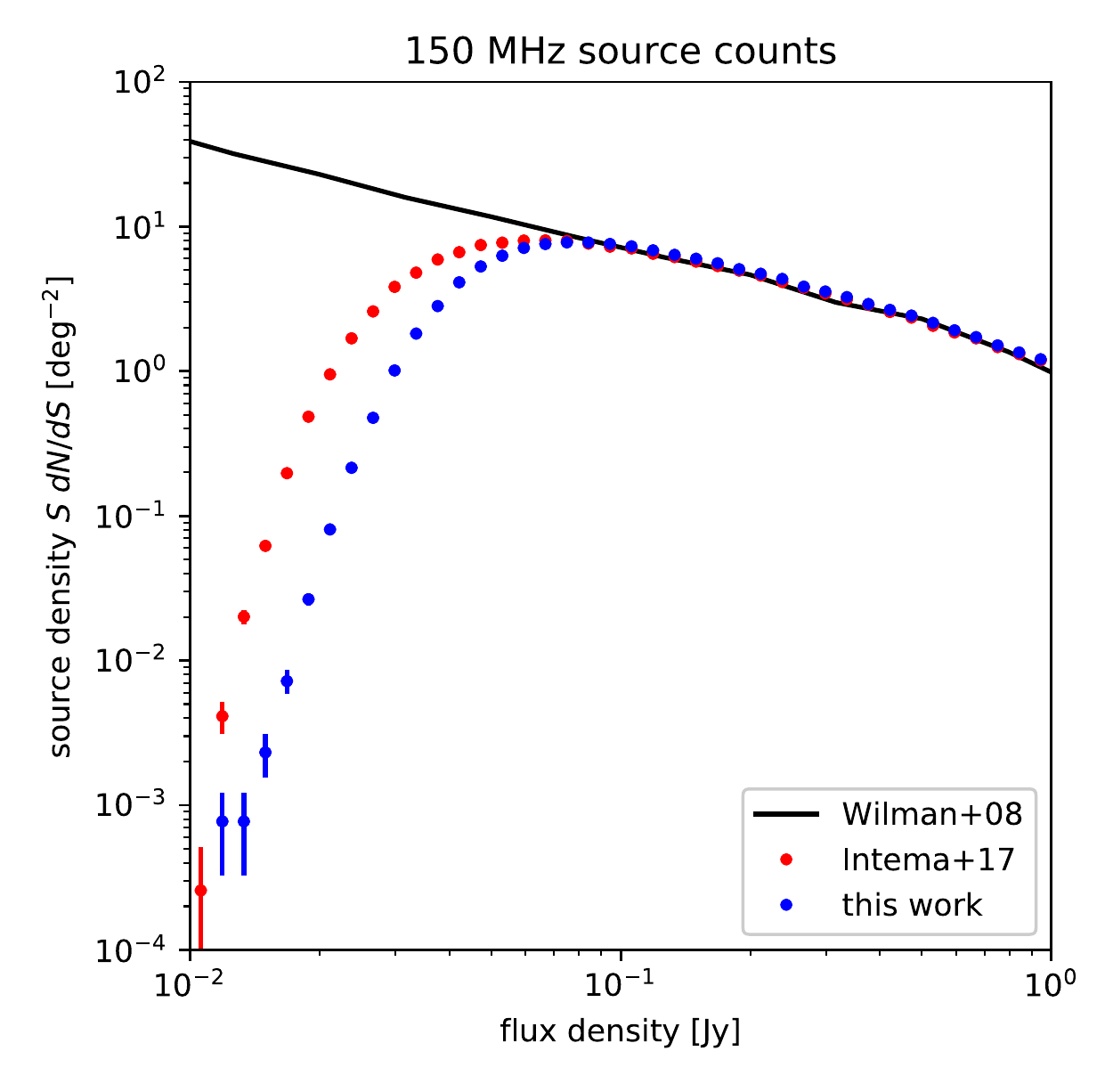}
\caption{Average number of sources per square degree as a function of flux density. The source counts of NVSS (above) and the reprocessed TGSS (below) are shown in blue dots, with Poissonian error bars. Red dots are from the original catalogue in \citet{Condon1998} and \citet{Intema2017}. Black line is the model source count at 1400 and 150 MHz derived from the SKADS simulation by \citet{Wilman2008}.}\label{fig:completness}
\end{figure}

We give here some basic information on the two derived narrow-band catalogues extracted from NVSS and TGSS images. Our NVSS catalogue (all full detections and lower limits) is of 1.3 million sources, compared to the 1.8 million of the official NVSS catalogue. The difference is most likely related to the different source extraction procedure. The original NVSS catalogue was produced by fitting Gaussians to all peaks brighter than 1.5~\mjybeam{} in the total intensity images \citep{Condon1998}. As evident from Fig.~\ref{fig:completness}, our catalogue misses several faint sources. This is understandable as most of them are lower limits selected with a $5\sigma_l$ cut, while the original NVSS catalogue cut is at $\sim3\sigma_{\rm avg}$. Furthermore, the original NVSS catalogue is a Gaussian list, while we took advantage of PyBDSF algorithm to obtain a list of sources by grouping nearby Gaussians, this is the origin of the slightly higher number of bright sources in our catalogue. We estimate our NVSS catalogue to be complete down to $\sim 5$~mJy.

The comparison with TGSS ADR1 published catalogue is harder, as in our case images have been reprocessed. From Fig.~\ref{fig:completness} we see that the higher noise of the reprocessed images reduce the number counts at low flux densities. However, the catalogue is complete down to $\sim100$~mJy. We also note that our catalogues are built in a special way so that sources detected only in one survey have a $5\sigma_l$ significance on that survey image, while full detections might have a lower significance (but always $\geq 4\sigma_l$) when considered on a single survey.

We also estimated the reliability (number of false positive) of our catalogues following \cite{Intema2017}. Since artefacts are both positive and negative in radio maps, we can estimate the number of false detection by running the source finder on inverted maps. This is done multiplying each pixel by $-1$ and re-running the source extractor. We find 7\,645 detections in NVSS inverted images, and 303 detections in the reprocessed TGSS inverted images. Only 1 full detection is found. This shows that the reliability of full detections is close to 100 per cent, while $\sim5$ per cent of TGSS upper limits are false detections and $<1$ per cent of NVSS lower limits are false detections.

\subsection{Cross-match completeness}
We investigate the ``cross-match completeness'' for the spectral index catalogue by looking at the number of detections in NVSS (TGSS) with a counterpart in the other survey. Using this definition, the cross-match completeness is much lower for NVSS than for TGSS, and is dependent on flux density.

We can estimate the catalogue cross-match \textit{in}completeness using the number of upper and lower limits. We divided NVSS (TGSS) detections in logarithmic flux density bins. Then, for each bin we estimated the ratio of lower (upper) limits to the total number of sources detected in that bin.

We plot this value in the bottom panels of Fig.~\ref{fig:spidxflux_distrib}. It is immediately clear that at low NVSS flux densities there are mostly lower limits. This is due to the relatively higher noise level of TGSS; only very steep sources are full detections and the catalogue is $<1$~per cent cross-match complete. Moving towards higher flux densities the catalogue become $>90$~per cent cross-match complete. Using TGSS as a reference survey is a more convenient choice. In this case the catalogue is always $>95$~per cent cross-match complete and $>99$~per cent cross-match complete above $S_{\rm TGSS} \gtrsim 0.1$~Jy.

We can also use upper (and lower) limits spectral index value to put stronger constraints on the cross-match completeness. Red and blue lines in the bottom right-hand panel of Fig.~\ref{fig:spidxflux_distrib} give an upper limit on the cross-match completeness for sources with spectral index $\alpha>-1.5$ and $\alpha>-1.0$. In the latter case we see that the catalogue cross-match completeness is $>99.9$~per cent in every flux density bin.

\begin{figure*}
\centering
\includegraphics[width=\textwidth]{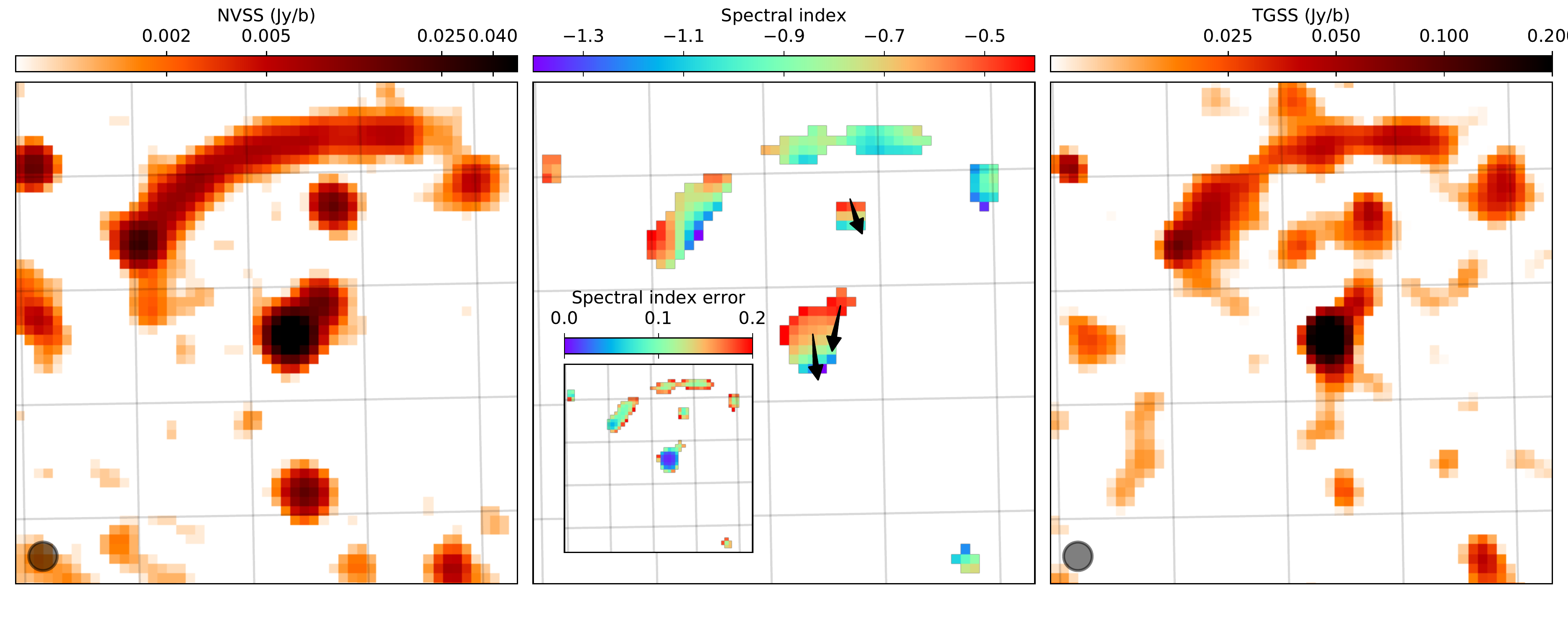}
\includegraphics[width=\textwidth]{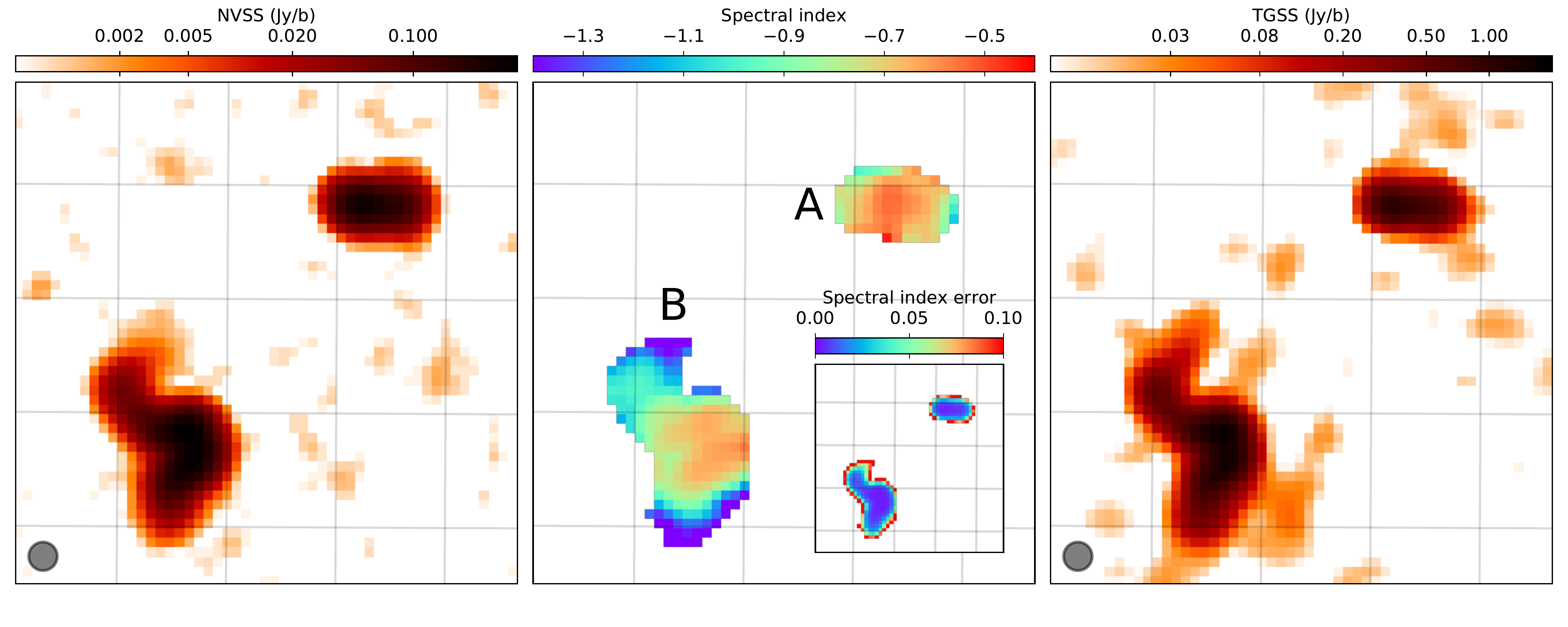}
\caption{Two examples of radio spectral index maps produced matching pixel-by-pixel NVSS and TGSS surface brightness maps. Left is the NVSS image, right the TGSS image, while in the centre we show the spectral index map and in the smaller panel the spectral index error map. Top panels: a galaxy cluster \citep[the ``Sausage'' cluster][]{vanWeeren2010a,Hoang2017}, where the spectral index gradients of the radio relic (the elongated structure on the north) and of three head-tail radio galaxies (with tails oriented as the three arrows) are easily visible. Bottom panels: two extended radio sources that can be classified thanks to the spectral index map into an FR\,I radio galaxy (source A) and a wide angle tail radio galaxy (source B).}\label{fig:spidxmap}
\end{figure*}

\section{Spectral index map}
\label{sec:map}

Together with the spectral index catalogue we produced a spectral index map that covers $\sim80$~per cent of the sky ($\delta>-40\deg$, where both NVSS and TGSS have available data). The map has been created by extracting a pixel by pixel spectral index value from the same sets of NVSS and TGSS images that were used to form the spectral index catalogue. From these images, we retained only those pixels with values above three times the local rms noise ($>3\sigma_l$) in both surveys. Pixels below this threshold were masked. For all non-masked pixels the spectral index and error were calculated by bootstrapping 1000~flux densities from both surveys assuming a Gaussian distribution with standard deviation equal to the local rms noise ($\sigma_l$). Pixels in the resulting spectral index maps contain the mean of the spectral index distributions as estimator, while the spectral index error maps contain the standard deviation of the bootstrapped distribution. We note that if an extended source is detected only in one of the two surveys due to different sensitivity to extended emission, the relative pixels were classified as limits and excluded from the published spectral index map.

\begin{figure}
\centering
\includegraphics[width=.5\textwidth]{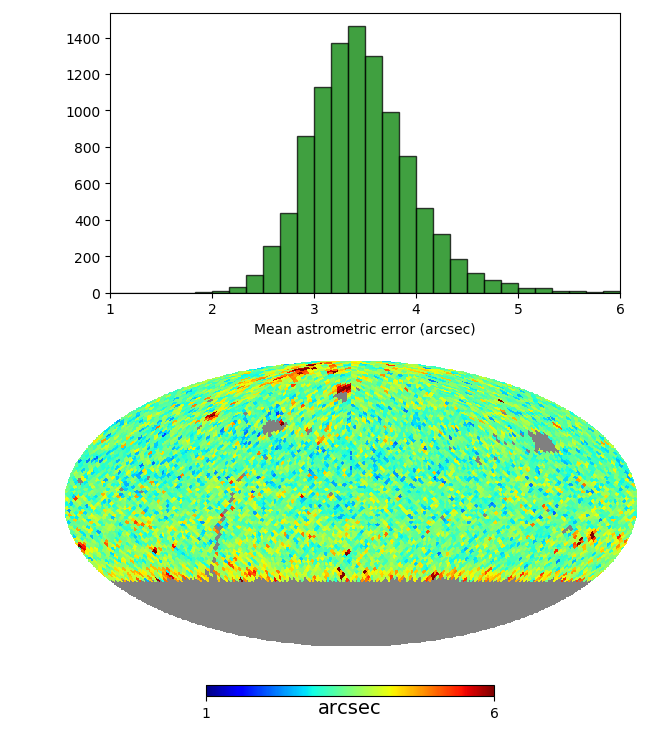}
\caption{To evaluate the uniformity of astrometric accuracy we pixellated the sky using a HEALPix scheme ($\rm NSIDE=32$, pixel area: 3.4 sq deg) and we evaluate the average astrometric error per pixel. Top panel: The distribution of the mean astrometric errors for full detections. Bottom panel: the distribution of the mean astrometric error on the sky. The errors are rather uniform apart from in the proximity of the north pole and at low declination.}\label{fig:astrometry_moll}
\end{figure}

An important aspect in producing spectral index maps is the astrometric accuracy. Low-frequency surveys usually have poor astrometric accuracy due to the effect of the ionosphere that can substantially displace sources from their real position. TGSS uses NVSS (and other surveys) to correct for astrometric error, this gives us some confidence that the two surveys are already aligned to the NVSS positions. We checked the distribution of the astrometric error using full detections in our catalogue. For both RA and Dec the mean difference between matched sources is $<1\arcsec$, with a standard deviation of $\approx3\arcsec$. We find a mean distance error of 3.5\arcsec (standard deviation: 2.7\arcsec). The mean astrometric error is rather uniform across the sky (see Fig.~\ref{fig:astrometry_moll}). Regions with slightly less accurate astrometry are present close to the north pole and at very low declination (where NVSS could not be used as astrometric reference).

Two examples from the NVSS-TGSS spectral index map are shown in Fig.~\ref{fig:spidxmap}. In both cases the spectral index map plays an important role in classifying the sources in the field. The top panel shows the well-studied merging galaxy cluster CIZA\,J2242.8+5301 \citep[the ``Sausage'' cluster;][]{vanWeeren2010a}. The elongated structure and spectral index gradient aids in the classification of the source as a radio relic. The three -- slightly extended -- sources below it also have gradients in the spectral index map, which suggests that they are head-tail radio galaxies. The orientation of the spectral gradient for the head-tail radio galaxies can be used to infer the direction of the galaxies' motion. These tentative classifications have been confirmed earlier using much deeper radio images \citep[e.g.,][]{Hoang2017}. The bottom panel shows another example on how the spectral index map can help with the classification. In the case of source A, both an FR\,I or and FR\,II radio galaxy could produce the morphological structure visible in the NVSS/TGSS survey. However, the steepening spectral index moving from the source centre to its edges point towards an FR\,I classification. This is confirmed with higher angular resolution FIRST images \citep{Becker1995}. Using the spectral index maps, even radio galaxies that are only mildly resolved can (in many cases) be classified as type FR\,I or FR\,II.

Full-sky interactive versions of the spectral index map and error map are available via \url{http://tgssadr.strw.leidenuniv.nl/spidx/} where they can be visually inspected in the web-browser through Aladin Lite \citep{Boch2014}, or imported as a Hierarchical Progressive Survey (HiPS) into a virtual observatory software such as Aladin \citep{Bonnarel2000}.

\section{Discussion}
\label{sec:discussion}

To illustrate the capabilities of this new catalogue, we used it to address two outstanding questions in extragalactic and galactic research. In Sec.~\ref{sec:spidxevol} we explore how and why the median spectral index changes as a function of flux density, and in Sec.~\ref{sec:gal} we investigate a previous claim of an excess of compact, steep-spectrum sources in the galactic plane.

\begin{figure}[t]
\centering
\includegraphics[width=.5\textwidth]{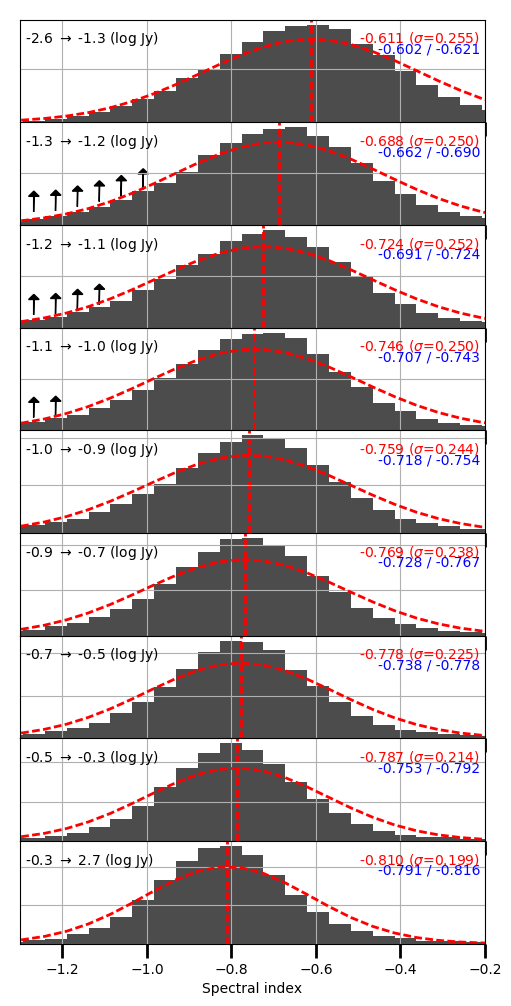}
\caption{From top to bottom, the distribution of spectral index of full detections for increasing TGSS flux density bins. The mean and standard deviation of fitted Gaussians are shown in red colour, and the mean and median value of each distribution are shown in blue colour. Arrows marks those flux density bins where there is a deficit of sources due to NVSS flux density limit (see Fig.~\ref{fig:spidxflux_distrib}), this bias the mean spectral index estimation towards a flatter values. In the top panel all bins are affected in different proportions, and the mean spectral index should be used as an upper limit. There is a trend for the average spectral index to steepening with increasing flux density.} \label{fig:spidxflux_evol}
\end{figure}

\begin{figure}
\centering
\includegraphics[width=.5\textwidth]{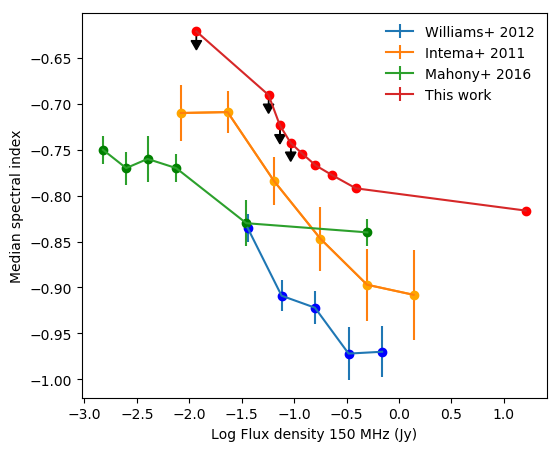}
\caption{Median spectral index variation per flux density bin. Data are taken from \citet{Intema2011}, \citet{Williams2012}, and \citet{mmp+16} and from Fig.~\ref{fig:spidxflux_evol} for this work. In all cases the spectral indexes of the fainter flux density bins are likely upper limits due to the limited sensitivity of high frequency images (upper limits are shown only for this work). Error bars for data points related to this work are too small to be visible.}\label{fig:spidx_evol_comp}
\end{figure}

\begin{figure}
\centering
\includegraphics[width=.5\textwidth]{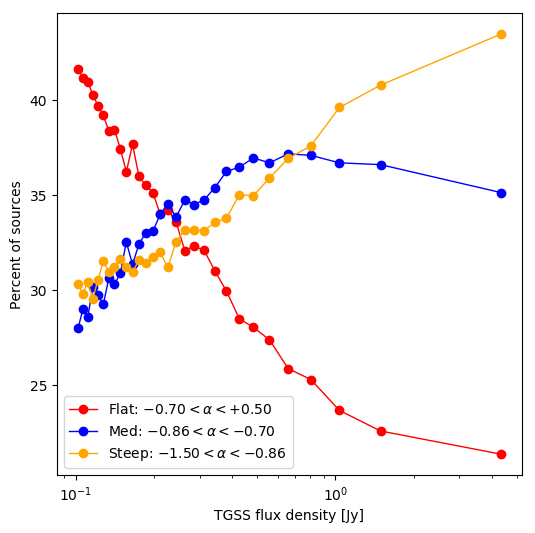}
\caption{Percentage of sources with flat ($-0.7<\alpha<+0.5$), medium ($-0.86<\alpha<-0.7$), and steep ($-1.5<\alpha<-0.86$) spectrum as a function of the TGSS flux density. Spectral index bins were chosen to have the same total number of sources. Flux density bins were also chosen to contain the same number of sources.}\label{fig:percent}
\end{figure}

\subsection{Spectral index as a function of flux density}\label{sec:spidxevol}

The radio sky is dominated by distant extragalactic sources \citep{condon1999}. For high flux densities at 1.4~GHz (10 mJy to 10 Jy), luminous radio galaxies and quasars dominate the source counts. With decreasing flux density ($\leq$1 mJy) there is an admixture of lower luminosity populations including star-forming galaxies, normal elliptical and spiral galaxies, and radio-quiet AGN. In the absence of redshift information, the spectral index and its variation with flux density can constrain evolutionary models for extragalactic radio sources and provide useful constraints on these changing source populations \citep{Laing1980,condon1984,CalistroRivera2017}. Extended, steep-spectrum radio sources ($\alpha<-0.5$) are most easily detected at low frequencies, and are likely lobe-dominated sources. Compact, flat spectrum radio sources ($\alpha>-0.5$) are most easily detected at high frequencies and are likely core-dominated AGN. A population of steeper spectrum, nuclear starbursts may emerge at flux densities lower than our survey detection thresholds \citep{pad2016}.

There have been a number of previous studies that have looked at how the median spectral index changes as a function of flux density. A possible trend has been seen for a flattening of the median spectral index with decreasing flux density \citep{ppw+06,isw+10,Intema2011} but the magnitude of the changes from one study to the next are not always consistent with each other. Some other studies see little or no evidence for a change in the spectral index \citep{kk86,sds+09}. The origin of these discrepancies are likely due to a combination of factors including the choice of frequencies, sample biases due to flux density thresholds, small sample sizes, and different angular resolution between surveys \citep[e.g.][]{zrrw03,mmp+16}. Compared to previous works, the catalogue presented here has a larger number of detections, fewer biases due to matched resolution images and matched source extraction, similar survey sensitivities, and a good sensitivity to a range of spectral indices.

With our large sample of spectral indices we can see a small but significant steepening of the average spectral index value with increasing source flux density. This is shown in Fig.~\ref{fig:spidxflux_evol} where sources were divided into TGSS-flux density bins of 55\,961 sources for which the spectral index distributions are plotted. As the average flux density increases, there is a trend toward steeper spectral indices. The upper four plots, with the lower flux density bins, are affected by the limited sensitivity of NVSS, and some steep-spectrum sources are missing. This can cause a bias towards a flatter average spectrum. Fig.~\ref{fig:spidx_evol_comp} plots these same data but now showing the trend of the average spectral index as a function of flux density, together with the results from several other previous studies.

Another way to look at this property is to divide all sources into three equally populated spectral index ranges, and plot the percentage of sources in each range as a function of flux density. This is shown in Fig.~\ref{fig:percent}. The percentage of steep-spectrum sources tends to increase with flux density from 30 to $>40$~per cent. Conversely, flat sources quickly start dominating the radio population at $S_{\rm 147\ MHz}<200$~mJy.

Qualitatively, a similar flattening trend is seen at both lower and higher frequency ranges by \citet{Cohen2003} and \citet{Tasse2006} in the range  of 74--1400 MHz, by \citet{zrrw03}, \citet{DeVries2001} and \citet{Owen2009} in the range of 325--1400 MHz, by \citet{Bondi2006} in the range of 610--1400 MHz, and by \citet{Prandoni2006} between 1.4 and 5 GHz. In the same frequency range we can compare our results with \citet{Intema2011}, \citet{Williams2012} and \citet{mmp+16}. For these cases the qualitative trend is similar, but the results are not compatible within each other (see Fig.~\ref{fig:spidx_evol_comp}). This discrepancy is higher than the possible systematic errors underlined in Sec.~\ref{sec:properties}. The discrepancy is likely due to combined errors in flux calibration of the order of $10-30$ per cent. Flux calibration seems to be still a relevant problem in low-frequency data reduction and it should be a warning for future radio surveys.

\subsubsection{Differences for compact and extended sources}\label{sec:compext}

\begin{figure}
\centering
\includegraphics[width=.5\textwidth]{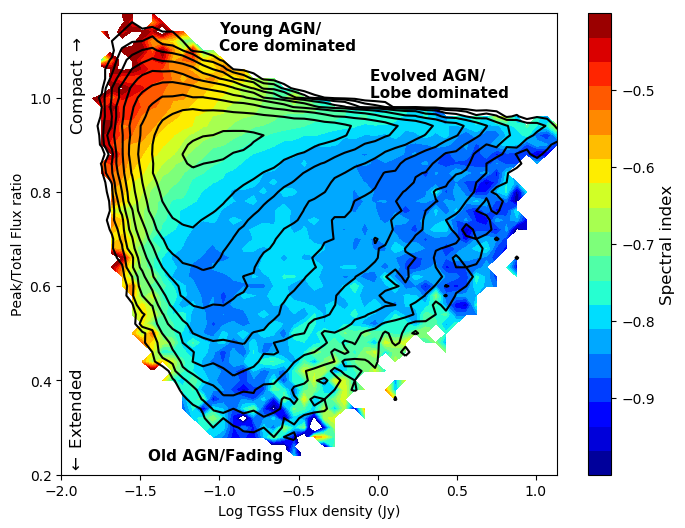}
\caption{Spectral indices of full detections are shown as a function of flux density and ``compactness''. As a proxy for compactness we used the average peak-to-total flux ratio in both surveys. A value of 1 means that the source is unresolved, smaller values implies an improper estimation of the total flux density from the source finder in a low signal-to-noise ratio regime. The range of parameters that is visualized was divided into $50\times50$ bins. The median spectral index value for each bin is colour-coded. Bins with less than five objects were blanked. Black contours show the source count per bin at 10, 20, 40, 80, 160, 320, 640, 1280, and 2560.}\label{fig:spidxflux_evol2}
\end{figure}

We further analyse the variation in average spectral index by selecting sources based on their compactness. As a proxy for source compactness we used the average source peak-to-total flux ratio from the two surveys. In Fig.~\ref{fig:spidxflux_evol2} we show the median spectral index value as a function of compactness and source flux density at 147 MHz.

The plot shows how compact/faint sources have in general a flatter spectral index than extended or bright sources. This can be attributed to the different part of radio galaxy emission they trace \citep[e.g.][]{Urry1995}. The core of the AGN is compact and its emission is typically sub-dominant compared to the emission from the lobes of a well-evolved radio galaxy. The radio spectrum of an AGN core is flat ($\alpha\simeq0.5$), while lobes are subject to adiabatic expansion, synchrotron and inverse Compton ageing that steepen their radio spectrum \citep{Blundell1999}. As a consequence, young/core-dominated radio galaxies will tend to cluster on the top-left part of Fig.~\ref{fig:spidxflux_evol2}, which is in fact predominantly populated by flat-spectrum sources. If the radio galaxy survives to the point of having large radio lobes, its luminosity and size increases, pushing the source towards the mid-right part of the plot. Here the spectrum is dominated by electrons of mixed ages in the radio lobes. Eventually, the AGN jets shut down, depriving the lobes from a source of energetic particles \citep{Shabala2008}. The lobes fade while continuing their expansion, and the source moves towards the bottom-left corner of the plot. Also this region appears to be populated by moderately steep-spectrum sources with a median spectral index value similar to that of evolved radio galaxies. This might appear unusual since many known remnant radio galaxies have steep spectra. Most of them were, in fact, discovered by selecting steep-spectrum sources \citep{Parma2007}. However, for dying radio galaxies selected by spectral curvature \citep{Murgia2011} or by morphology \citep{deGasperin2014a,bgmv16}, moderate spectral indices have often been observed.

The simple interpretation above can, in a qualitative sense, explain some of the properties of Fig.~\ref{fig:spidxflux_evol2}, and provides an explanation for the (average) spectral index dependency on source flux density noticed in the previous section. However, our sample covers a large redshift range, resulting in some degeneracy between source size and distance, i.e. compact source could be intrinsically compact or just at high redshift. Furthermore, we describe the properties of the sample assuming that radio galaxies are the only population present, while in certain regions of the plot this is probably not true. For example, radio loud quasars will likely be an important fraction of compact and bright sources that flatten the top right region of the plot. Finally, we note that theoretical modelling of classical double-double radio galaxies \citep[the dominant population of our sample;][]{Wilman2008} predicts both the power -- spectral index and the linear size -- spectral index trends we find \citep{Blundell1999}.

\subsection{An excess of steep-spectrum sources on the galactic plane}
\label{sec:gal}

The steep-spectrum tail ($\alpha\lesssim-1.5$) of the spectral index distribution in Figs.~\ref{fig:spidx_kde} and \ref{fig:spidxflux_distrib} is home to several rare, exotic source populations. Extended steep-spectrum sources include radio haloes and relics in merging galaxy clusters \citep{Feretti2012}, and steep-spectrum lobes that are argued to be the remnants from previously active phases of radio galaxies and AGN \citep[e.g][]{sj09,smos12}. Selecting steep-spectrum radio sources is also an efficient technique for finding luminous, high-redshift ($z>2$) radio galaxies (HzRGs). HzRGs are thought to be distant analogues of the massive elliptical galaxies in nearby clusters, and as such they can be used to trace structure formation (i.e. proto-clusters) in the early universe \citep{isw+10,Afonso2011,sbw+14}.

Within our galaxy the only known sources of steep-spectrum radio emission are pulsars. From 100 MHz to a few GHz the majority of pulsars have single power-law spectra with average slopes of $\alpha= -1.8\pm0.2$ \citep{mkk+00}. While pulsars are defined by their pulsed, periodic emission, they can also be detected in radio images as time-averaged continuum point sources. Searching through the TGSS, \cite{fjmi16} detected continuum radio emission from nearly 300 {\it known} pulsars. A similar search using NVSS yielded 79 known pulsars \citep{Kaplan1998}, while a later search \cite{Han1999} identified 97 known pulsars. Once corrected for observational selection effects, the intrinsic average pulsar slope is $\alpha=-1.4\pm{1.0}$ \citep{blv13}. These values are in line with recent samples selected at low frequencies \citep[$\sim 100-200$ MHz;][]{bkk+15,fjmi16,bmj+16,mkb+17}. Polarisation can be another useful discriminant since pulsar profiles are often strongly polarised in linear and circular polarisation  \citep{Han2009}. In the image plane the fractional polarisation will be phase-averaged but it is still strong enough in some cases to identify pulsar candidates \citep[e.g.][]{Navarro1995}.

In a search for HzRGs, \citet{bbrm00} created a data base of 143\,000 spectral indices from the WENSS (325 MHz) and NVSS (1400 MHz) catalogues. When they selected all sources with $\alpha<-1.6$ and examined the distribution of as a function of galactic latitude $b$ they saw an excess in the fraction of steep-spectrum sources near the galactic plane ($\vert{b}\vert<15^\circ$) and suggested that these may be previously unknown pulsars. \citet{dpf11} searched for pulsations at 1.4~GHz using a sub-sample of 24~unresolved candidates with a polarised intensity (at 1.4 GHz) of $\geq5$~per cent. No pulsars were found.


\subsubsection{Sample selection and systematic errors}

To confirm the excess claimed by \citet{bbrm00} and to test whether pulsars are a plausible explanation for this excess, we first need to define a suitable sample from the spectral index catalogue and ensure there are no systematic errors. When searching for {\it unknown} pulsars in the image plane the two most important criteria are spectral index and compactness.

To obtain improved constraints on source positions and compactness, we created a new source sample by cross-matching the spectral index catalogue with a special 5-sigma version of the full-resolution (25\arcsec) TGSS catalogue. Note that the public TGSS ADR1 catalogue is a more conservative 7-sigma catalogue. Sources without a counterpart were dropped from the sample, which removed most of the lower limits. The resulting sample consisted of 530\,391 sources (5\,574 upper limits and 71\,116 lower limits). The remaining lower limits are sources undetected in the 45\arcsec{} resolution TGSS images but are present in the more complete 25\arcsec{} TGSS catalogue.

Next, we removed {\it known} pulsars from the sample. This is aided by the improved positional (astrometric) accuracy as described above, especially in the galactic plane where source confusion is worse than in other parts of the sky. In our sample we found a total of 128 known pulsars. This number is lower than what was found by \cite{fjmi16}, due to the higher noise of the low-resolution TGSS maps that were used to determine the spectral indices.

For our sample we calculated the \emph{source compactness} $C$ based on source properties originating from the 25" TGSS catalogue. For this, we used an improved version of Eq.~3 in \cite{Intema2017} that depends on the source total flux ($S_{\rm total}$), peak flux ($S_{\rm peak}$), as well as on the local rms noise ($\sigma_l$):
\begin{equation}
C = \frac{1.071 + 2 \cdot \sqrt{0.038^2 + 0.39^2 \cdot \left(S_{\rm peak}/\sigma_l \right)^{-1.3}}}
{ \left(S_{\rm total}/S_{\rm peak}\right)}.
\end{equation}
The value of $C$ is expected to be $\geq 1$ for sources that are point-like in the $25\arcsec$ TGSS catalogue, and $<1$ for extended sources.

\begin{figure}
\centering
\includegraphics[width=.5\textwidth]{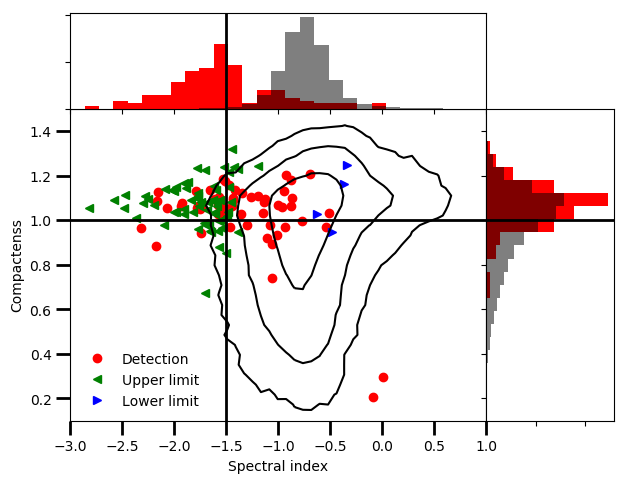}
\caption{Contours show the source density (excluding upper and lower limits) in a compactness versus spectral index plot. Sources with compactness $> 1.0$ can be considered point sources. Markers shows the location of known pulsars detected in our refined catalogue. They tend to cluster in the compact/steep-spectrum region of the distribution. The two sources with low compactness and flat spectrum are detection of supernova remnant surrounding the pulsar \citep[SNR G54.1+0.3 and SNR G21.6-0.8;][]{Green2014a}. Histograms are normalised by area and shows the distribution of matched pulsars (red) and all the other sources (grey). Solid line shows the location of the cuts used to produced Fig.~\ref{fig:pulsar_hist_lat}.}\label{fig:pulsar_selection}
\end{figure}

In Fig. \ref{fig:pulsar_selection} we plot the compactness parameter ($C$) versus spectral index for this sample. The distribution of the source properties of the sample as a whole is much like what has been discussed previously in Secs.~\ref{sec:properties} and \ref{sec:spidxevol}. For comparison, we plot the properties of the known pulsars as identified in the spectral index catalogue. As we noted above, the spectral index distribution of pulsar spectral indices skews to negative values. Only 1~per cent of all full detections and upper-limits have $\alpha<-1.5$, while 61~per cent of all known pulsars have such steep values. Pulsars are also compact with the majority (77 per cent) having $C>1$ and cluster in the upper left of Fig. \ref{fig:pulsar_selection}.

Although physically compact, pulsars may still have measured values of $C$ below unity. Both extended emission and interstellar scattering can create outliers in the $S_{\rm total}/S_{\rm peak}$ ratio and hence the compactness parameter. Extended emission from a co-located supernova remnant, pulsar wind nebula, or nearby \Hii{} region can confuse the flux density measurements in the image plane. Furthermore, interstellar scattering of the radio waves from pulsars can cause both spectral and temporal variations \citep{Cordes1985}, and the subsequent amplitude variations that can occur on a time-scale of the interferometer integration time complicate the continuum imaging process. These variations can cause persistent image artefacts at the pulsar location, which in turn leads to larger errors in the source extraction and the determination of the compactness \citep[e.g.][]{fjmi16}.

\begin{figure}
\centering
\includegraphics[width=.5\textwidth]{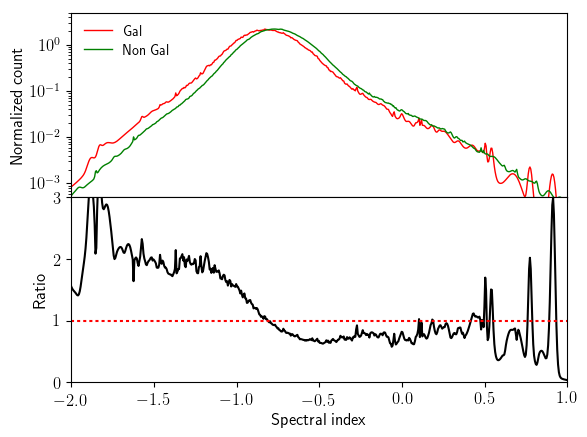}
\includegraphics[width=.5\textwidth]{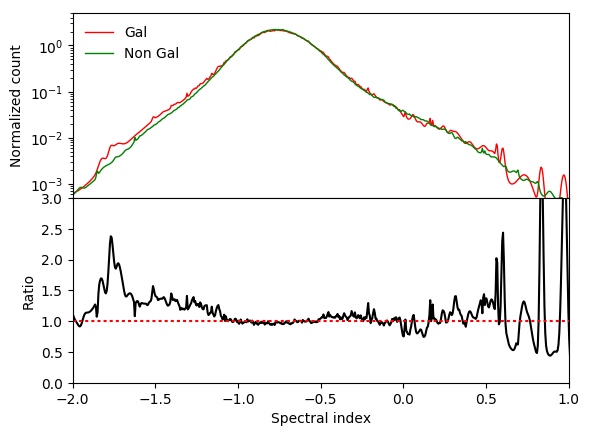}
\caption{Normalised source count before (top) and after (bottom) the correction of the spectral index for sources on the galactic plane. The second paired panels show the ratio between the low- and high-latitude distributions. A mild excess of steep-spectrum sources is still present on the low-latitude sample after the correction.}\label{fig:bias}
\end{figure}

As a final step, we grouped our sources into low galactic latitude ($\vert{b}\vert<5^\circ$) and high galactic latitude ($\vert{b}\vert>15^\circ$). The median spectral index value for the low-latitude sources is $\alpha=-0.82\pm0.04$, while for the high-latitude sources is $\alpha=-0.75\pm0.04$. These values are significantly different\footnote{To verify that this is an anomalous deviation we pixellate the sky using and HEALPix tessellation ($\rm NSIDE=4$, corresponding to a pixel area of $15\deg \times 15\deg$) and retaining only sources outside the galactic plane ($-15>$ galactic latitude $>+15$). The standard deviation of the median spectral index per pixel is found to be 0.03.} and pointed us towards a possible systematic effect in one or both of the surveys. We found that the spectral index distribution of galactic sources was systematically offset towards steeper spectra compared to the distribution of the extragalactic ones. This effect is apparent in Fig.~\ref{fig:bias} (top), where we show the distribution of the high and low-latitude samples as a function of spectral index. The ratio of the low and high-latitude samples show a systematic trend with an excess of steep-spectrum sources and a deficit of flat and positive spectral index sources on the galactic plane.

This offset and systematic trend can be readily explained if there is a combined (average) flux density error of 14.6 per cent in the two surveys for fields on the galactic plane. To test this possibility, we compared the median flux density values for full detections with $\rm{s2n}>20$ within 5\deg{} and further than 10\deg{} from the galactic plane. We noted that galactic TGSS flux densities are higher by $\sim 21$~per cent, while NVSS sources are on average $5$~per cent brighter on the galactic plane. The combined effect brings to an error of $\sim 16$~per cent, similar to what we expected. Part of the effect might come from the different populations present inside and outside the galactic plane; however at TGSS/NVSS flux density limits the bulk of the radio population is expected to be extragalactic. Calibration errors can explain some of the excess claimed by \citet{bbrm00}. In the rest of this section, to account for this discrepancy we rescaled the spectral index of galactic sources by $+0.06$. In Fig.~\ref{fig:bias} (bottom) we show effects of this simple scaling that brings the median spectral index of the high and low-latitudes samples into agreement and flattens the trend line.

\begin{figure}
\centering
\includegraphics[width=.5\textwidth]{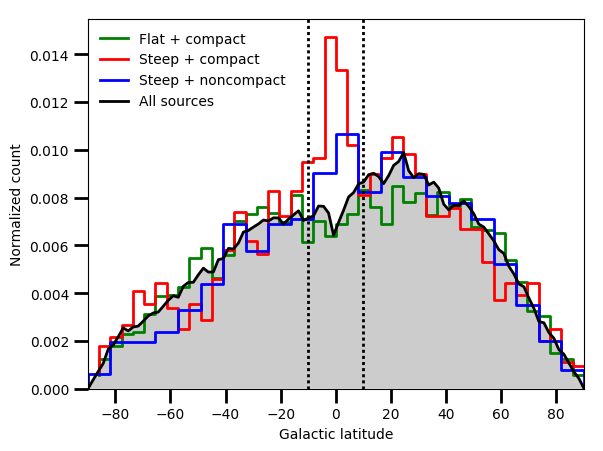}
\caption{Galactic latitude distribution of all sources and upper limits (grey) and of sources with Compactness $C>1$ and spectral index $\alpha<-1.3$ (red) and $\alpha>0$ (green). Known pulsars were removed from this sample. The two histograms were re-normalized to have a matched area in the region outside galactic latitude $10\deg$. The skewed shape of the all-source histogram is due to the non-complete coverage of the sky, the small deficit of sources at galactic latitude $<5\deg$ is due to a local increase in noise on the galactic plane. An excess of compact, steep-spectrum sources is visible at low galactic latitude.}\label{fig:pulsar_hist_lat}
\end{figure}

After the bias is removed, there is still a slight excess in the number of steep-spectrum sources at low galactic latitude (Fig.~\ref{fig:bias}). To investigate this further, we plot the distribution of our sources as a function of galactic latitude in Fig. \ref{fig:pulsar_hist_lat}. The distribution of {\it all}-source counts shows a small but expected dip in the galactic plane, where there is a strong rise in the system temperature due to large-scale galactic synchrotron emission, and hence a drop in the number of detectable sources. If we apply a spectral index cut-off $\alpha<-1.5$, we see an excess of steep-spectrum sources in the region of the galactic plane ($\vert{b}\vert\leq10^\circ$).  The same excess is not seen if we apply other spectral index cuts to the source list (e.g. $\alpha>0$). Compactness also appears to be a significant factor in characterizing the excess, although the excess seems not to be due to only compact sources (see Fig.~\ref{fig:pulsar_hist_lat}). For instance, if we select $\alpha<-1.5$ and $C>1$, we find an excess of $\sim$86 sources or 28~per cent, while for steep-spectrum and non-compact sources ($C<1$) the excess on the galactic plane remains but goes down to 16~per cent.

\subsubsection{Possible origins of galactic plane excess}

Our result confirms the steep-spectrum excess in the galactic plane first seen by \citet{bbrm00}. Possible support for this excess can also be seen in the GLEAM survey \citep{Hurley-Walker2017}. Although GLEAM has not released data for the entire galactic plane (i.e. $\vert{b}\vert>10^\circ$), a plot similar to Fig. \ref{fig:pulsar_hist_lat} shows the number of steep-spectrum GLEAM sources rising towards decreasing galactic latitudes.

The simplest explanation for an excess of steep-spectrum/compact sources is that they originate from undiscovered radio pulsars. Normal, rotation-powered pulsars have characteristic periods of 500~ms, ages of 10$^7$~yrs, and dipole magnetic fields of $10^{12}$~G, and are concentrated in the plane of the galaxy from which they are born \citep{fk06}. The recycled or millisecond pulsars have characteristic periods of 10~ms, ages of 10$^9$~yrs and dipole magnetic fields of $10^{8}$~G, but they have much larger scale heights \citep{dunc08}. The latitude distribution of the sources comprising the excess suggests that the bulk of the sources are normal pulsars with longer periods than recycled pulsars. The number of prospective pulsars is not small. The excess, if attributed entirely to pulsars, would be a good fraction ($\sim5$~per cent) of the $\sim$2000 pulsars currently known in our Galaxy. This begs the question of how so many pulsars could have been missed in past searches, and what strategies might be needed in order to identify this putative population.

Current state-of-the-art pulsation searches at 1.4~GHz are sensitive to longer-period pulsars with mean flux densities of $0.05-0.15$~mJy at low-latitudes ($\vert{b}\vert<3.5^\circ$), and $0.2-0.25$~mJy at mid-latitudes ($\vert{b}\vert<15^\circ$) \citep{kjs+10}. These search limits are all well below the sensitivity limit of NVSS and so the majority of the excess should be detectable if they are pulsars. There are compact TGSS-only detections on the galactic plane whose extrapolated flux density at 1.4 GHz would render pulsations undetectable, but these are not numerous ($<1$~per cent). Other explanations for unsuccessful pulsation searches, despite there being detectable radio continuum, include short-period, eccentric binaries missed due to reduced signal-to-noise ratio in acceleration searches \citep{ncb+15}, or pulsars with unpulsed emission because their magnetic dipoles are aligned with their rotation axes \citep{pl85}. Neither of these exotic explanations are likely to account for the bulk of the excess population, however. Enhanced interstellar scattering may offer a better alternative. The lines of sight through the Galaxy encounter turbulent ionised gas that broaden the observed pulsar width. Pulse scatter times $\tau_{\rm scat}$ increase with the pulsar ionised gas column density, also called dispersion measure (DM). At larger DM values the observed $\tau_{scat}$ can vary by two orders of magnitude or more \citep{bcc+04}. If the excess sources are pulsars with larger than average scattering, then pulsation searches above 1.4 GHz are needed, since temporal scattering is a strong function of frequency ($\tau_{\rm scat}\propto\nu^{-4}$).


Other non-pulsar explanations might account for the excess. Radio transients or strong variables can produce outliers in the positive or negative tails of the spectral index distribution. A transient source appearing in TGSS but not in the NVSS would be identified by its steep spectrum. Transients are a negligible source of false positives. From a comparison of TGSS and GLEAM the transient surface density at 150 MHz is estimated to be 6.2$\times 10^{-5}$ deg$^{-2}$ \citep{mkc+17}. Similar low transient rates are found at 1.4 GHz \citep{gop+06}, and there is no evidence that transient rates increase toward the galactic plane at these sensitivities \citep{wbc+13}.

Strong variability can in principle be an important contaminant. There is a 16-year difference between the two surveys. The mean epoch for the TGSS catalogue is 2011 January 18, while the NVSS was observed around 1995. All radio sources vary to some degree. At metre wavelengths, the variations are dominated by propagation phenomena, mostly refractive effects. At centimetre wavelengths, variations are dominated by intrinsic effects, i.e. changes in the black hole-accretion disc environment \citep{dan89}. Strong variability (i.e. $\sim40$~per cent) can produce changes in the mean spectral index of order $\pm$0.15. This amounts to a simple error term when studying the peak of the radio source population in Fig. \ref{fig:spidx_kde}, but the outlying tails of the spectral index distribution are more susceptible. Even though only a fraction of radio sources may show this level of variability, the smaller number of {\it intrinsically} negative (and positive) steep sources will be contaminated by the more numerous flatter spectral index sources scattering into the tail.

At 1.4~GHz there is no indication that strong variability is contributing to the steep-spectrum excess. The radio sky is remarkably quiet at centimetre wavelengths. Given the sensitivity of the NVSS, strong variables are present at levels of one per cent \citep{mhb+16,hdb+16}. Moreover, when strong variables are identified in the galactic plane, they have flat or positive spectral indices \citep{bhwp10}. Strongly variable, centimetre-wavelength radio sources are not numerous enough nor do they have the right spectral index distribution to affect the steep-spectrum tail. Less is known about the variability of the metre-wavelength sky on these time-scales, but we can test for the effects of variability using the spectral index data from the GLEAM sample (see Sec.~\ref{sec:properties}). Source variability will not affect the GLEAM sample since the multi-frequency data are taken at the same epoch. The multi-point GLEAM spectral indices for the $\alpha<-1.5$ sample are somewhat steeper relative to our two-point values at decreasing flux densities (Fig.~\ref{fig:delta}). This suggests that we have {\it underestimated} the number of steep-spectrum sources, and the effect is the opposite of what would be expected if the steep-spectrum tail of the distributions was contaminated by strong variables.

\section{Conclusions}\label{sec:conclude}

We have presented the largest radio spectral index catalogue assembled to date. The data were extracted from the NVSS (1400~MHz) and a re-imaged version of the TGSS (147~MHz). In contrast to previous studies we did not cross-match published catalogues, but instead we detected sources on images made with the same gridding and resolution combining overlapping regions of emission. This procedure overcomes some systematic errors and allows for a lower (combined) detection threshold for marginal detections in both surveys. The final spectral index catalogue includes  503\,647 full detections and 851\,845 upper/lower limits. We also provide the community with a pixel-by-pixel radio spectral index map of 80~per cent of the sky at 45\arcsec{} resolution. The catalogue and spectral index map are available online via \url{http://tgssadr.strw.leidenuniv.nl/spidx}.

We used the new catalogue to address two important questions. In the first case we looked at how and why the median spectral index changes as a function of flux density. We found clear indication of spectral steepening with increasing flux density. We also found that this trend is due to a large fraction of compact, flat-spectrum ($\alpha \simeq -0.5$) sources that populate the faint end of the flux density distribution. We argue that these characteristics are driven by the presence of core-dominated and young AGNs. Moderate redshift, well-evolved sources make up the bulk of the bright portion of this phase space. Their spectral index is steeper ($\alpha \simeq -0.8$) since their emission is dominated by their extended lobes.

Secondly, we used the catalogue to investigate an earlier claim of an excess of steep-spectrum sources on the galactic plane. After removing the known pulsars and correcting for a systematic spectral index offset in the galactic plane, we confirm an excess. The properties of the excess are consistent with normal non-recycled pulsars. We suggest that this pulsar population may have been missed by past pulsation searches due to larger than average scattering along the line of sight.

\subsection{Future Work}

We have made our radio spectral index catalogue and our radio spectral index map publicly available in anticipation that it will be used for other research projects. For example, we note that the {\it isotropic} component of the compact, steep-spectrum sources in Fig.~\ref{fig:pulsar_hist_lat} are excellent candidates for HzRGs. Actually, one of the prime motivation for locating ultra-steep-spectrum radio sources is finding more HzRGs in part because HzRGs trace proto-clusters, and in the case of the most distant objects, they could be used to measure \Hi{} absorption in the early universe \citep{Carilli2004,isw+10,mmp+16,sbw+14}. Our sample consists of 4130 compact ($C>1$), steep-spectrum ($\alpha<-1.3$) sources. Ignoring the galactic plane excess, we estimate the areal density of isotropic component is 0.1 deg$^{-2}$. While our areal density is comparable to past surveys \citep[e.g.][]{bbrm00}, the number of candidate HzRGs is more than an order of magnitude larger, owing to the wide sky area covered and the lower flux density threshold.

Identifying genuine HzRGs will require optical/infrared identifications with spectroscopic follow-up \citep{bbv+07}. Likewise, much could learned by cross-matching the {\it entire} radio spectral index catalogue with the Sloan Digital Sky Survey (SDSS). The better resolution of TGSS relative to NVSS should allow for more efficient cross-matching of a large spectral index sample. Morphological information from SDSS could be used to test our initial conclusions about compact and extended source populations from Sec.~\ref{sec:compext} and Figs. \ref{fig:spidxflux_evol}--\ref{fig:spidxflux_evol2}. For a smaller radio sample, \citet{ki08} used SDSS redshifts to investigate how the radio luminosity and the radio spectral slope depended on optical quantities such as colour, magnitude and redshift, and they used this sample to test for effects of evolution and environment. At higher energies, our radio spectral index catalogue could be used to search for blazar candidates counterparts for the unidentified gamma-ray sources \citep{bbv+07}, or to investigate the radio properties of X-ray-selected AGN.

Within the galactic plane, the radio spectral index map can be used to identify compact \Hii{} regions from their flat or inverted (i.e. positive) spectral indices. Extended, non-thermal radio sources in the plane with $\alpha\sim-0.5$ are likely to be supernova remnants. Young remnants with ages of a few hundred years are missing from existing samples \citep{dag05}. The spectral index map can be used efficiently to identify small diameter, non-thermal sources in the galactic plane worthy of further follow-up \citep[e.g.][]{rbg+08}. Finally we note that some of the high galactic latitude sources with the steepest spectral indices are likely to be recycled millisecond pulsars. Cross-matching of such sources with optical light curves or X-ray/gamma-ray emission could identify promising candidates worthy of follow-up pulsation searches.

\section*{Acknowledgements}
We thank Jeremy Harwood for the important help in the interpretation of Fig.12 in terms of radio galaxy evolution and for other caveats pointed out in Sec. 4.1.1.
FdG is supported by the VENI research programme with project number 1808, which is financed by the Netherlands Organisation for Scientific Research (NWO).
The GMRT is run by the National Centre for Radio Astrophysics of the Tata Institute of Fundamental Research.
The VLA is run by the National Radio Astronomy Observatory, a facility of the National Science Foundation operated under cooperative agreement by Associated Universities, Inc.
This research has made use of SAOImage DS9, developed by Smithsonian Astrophysical Observatory.
This research has made use of NASA's Astrophysics Data System.

\bibliographystyle{mn2e}
\bibliography{papers-spidxskymap}
\bsp

\label{lastpage}

\end{document}